\documentclass[preprint,showpacs,epsfig,aps]{revtex4}
\usepackage{graphicx}
\usepackage{epsfig}
\usepackage{verbatim}

\begin{document}
\title{Alternate Derivation of Ginocchio--Haxton relation $[(2j+3)/6]$}

\author{L. Zamick and A. Escuderos}

\affiliation{Department of Physics and Astronomy, Rutgers University,
Piscataway, New Jersey USA 08854}


\begin{abstract}
We address the problem, previously considered by Ginocchio and Haxton (G--H),
of the number of states for three identical particles in a single $j$-shell
with angular momentum $J=j$. G--H solved this problem in the context of the
quantum Hall effect. We address it in a more direct way. We also consider the
case $J=j+1$ to show that our method is more general, and we show how to take
care of added complications for a system of five identical particles.
\end{abstract}
\maketitle

We here address  a problem originally solved by Ginocchio and Haxton 
(G--H)~\cite{gh93}, the number of total angular momentum $J=0$ states for four 
fermions in a single $j$-shell. Amusingly, they published their results in a 
paper with the title ``The Fractional Quantum Hall Effect and the Rotation
Group". This is not exactly the first place one would look for the solution to 
this probem. Also the work is published in a not easily accessible 
compilation~\cite{gh93}.

    We are going to immediately switch this to a problem of three identical 
fermions (e.g. neutrons, electrons, etc.) in a single $j$-shell. It is easy to 
show that the number of $J=j$ states for three fermions is the same as the 
number of $J=0$ states for four. The method we apply does not require us to 
construct the detailed wave functions.

To obtain the number of $J=j$ states for three fermions we follow the advice of
Talmi~\cite{t93}, who states that first one should calculate the number of 
states with $M=(j+1)$; this is equal to the number of states with $J>j$. Then, 
we subtract from this the number of states with $M=j$. However there are no 
further details given in his book. We here give our version.

Let us consider 3 neutrons in the $j$-shell, each in a $(j,M_i)$ state. We can
form states with total angular momentum $J$ and $M=M_1+M_2+M_3$, where 
$M_1>M_2>M_3$. Now we consider the states with $J=j+1$ and
$M=j+1$. These basis states are labelled ($M_1,M_2,M_3$). From the latter we 
then create states with $M=j$ of the form ($M_1,M_2,M_3-1$). These states all 
exist because the smallest possible value of $M_3$ is $(j+1)-j-(j-1)=-j+2$. 
(Only if $M_3$ were $-j$ would we have to worry).

At this point we still have the number of states with total $J$ greater than 
$j$. If there are any additional states with $M=j$, that number will be the 
number of states with $J=j$, since obviously states with $J<j$ cannot have 
$M=j$. The additional states with $J=j$ which cannot be obtained by lowering 
$M_3$ are listed in Table~\ref{tab:M.eq.j} for various examples.
        
\begin{table}[!h]
\begin{tabular}{c|c|c|rc||rc|c|c|r}    
\makebox[.5in]{$j$} & \makebox[.5in]{$M_1$} & \makebox[.5in]{$M_2$} & 
\makebox[.5in]{$M_3$} & \hspace{1cm} & \hspace{1cm} &
\makebox[.5in]{$j$} & \makebox[.5in]{$M_1$} & \makebox[.5in]{$M_2$} & 
\makebox[.5in]{$M_3$} \\
\cline{1-4} \cline{7-10} 
7/2 & 7/2 & 1/2 & $-1/2$ & & & 13/2 & 13/2 & 1/2 & $-1/2$ \\ 
\cline{1-4}
9/2 & 9/2 & 1/2 & $-1/2$ & & & & 9/2 & 3/2 & 1/2 \\
\cline{7-10}
 & 5/2 & 3/2 & 1/2 & & & 15/2 & 15/2 & 1/2 & $-1/2$ \\
\cline{1-4}
11/2 & 11/2 & 1/2 & $-1/2$ & & & & 11/2 & 3/2 & 1/2 \\
 & 7/2 & 3/2 & 1/2 & & & & 7/2 & 5/2 & 3/2 \\
\end{tabular}
\caption{\footnotesize States with $M=j$ which cannot be obtained by lowering 
$M_3$ in states with $M=j+1$.}
\label{tab:M.eq.j}
\end{table}

From these examples, we see that $M_1+M_2+M_3=j$ (obvious) and that 
$M_3=M_2-1$. We can understand the latter condition by noting that if, for 
$M=j$ we apply the raising operator on $M_3$, we would get a state of the form 
($M_1,M_2,M_2$), and this is forbidden by the Pauli Principle.

Now let $n$ be an integer ranging from zero to $N-1$, where $N$ is the number 
of states with $J=j$. We see that $M_1$ takes the general form $M_1=j-2n$; then
$M_2=1/2+n$, with $n=0,1,2,\cdots,N-1$, as stated above. Furthermore, $M_1$ is 
bigger than $M_2$; thus, we have
\begin{equation}
j-2n> \frac{1}{2}+n,
\end{equation}
which leads to 
\begin{equation}
n < \frac{2j-1}{6}\, .
\end{equation}
Since $N$ is an integer and $N=n_{\rm max}+1$, we obtain our final result
\begin{equation}
N < \frac{2j+5}{6},
\end{equation}
which means that $N$ is equal to the largest integer that is less than 
$(2j+5)/6$.

It does not immediately look the same as the Ginocchio--Haxton 
relation~\cite{gh93}, but can easily be shown to be the same. Note that $2j+5$ 
is even. Thus, $(2j+5)/6$ is either $I,I+1/3$ or $I+2/3$, where $I$ is an 
integer.
            
If the answer is $I$, then $N=I-1$
\begin{equation}
\left[\frac{2j+3}{6}\right]= \left[I-\frac{1}{3}\right]= I-1.
\end{equation}
If $I+1/3$ is the answer, then $N=I$
\begin{equation}
\left[\frac{2j+3}{6}\right]=[I]=I.
\end{equation}
If $I+2/3$ is the answer, then $N=I$ and $[I+1/3]$ also is equal to $I$.

As Talmi notes~\cite{t93}, of the $N$ states, one will have seniority $v=1$ and
$N-1$ will have seniority $v=3$. The quantity $N-1$ is equal to $[(2j-3)/6]$.  
     
Note that although we use the states $j-2n,n+1/2,n-1/2$ to count the number of 
$J=j$ states, it does not mean that these are $J=j$ states. In fact, they are
not, even after antisymmetrization. This is a novel feature of this work. We
use the basis ($M_1, M_2, M_3$) for counting purposes, but we do not need to
construct the correct $J=j$ wave functions.

\begin{table}[!hb]
\begin{tabular}{c|c|rc||rc|c|rr}
\multicolumn{8}{c}{$j=15/2$ ($M=M_1+M_2+M_3$, $M_1>M_2>M_3$)} \\
\hline
\multicolumn{3}{c}{$M=j+1=17/2$} & \hspace{1cm} & \hspace{1cm} & 
\multicolumn{3}{c}{$M=j=15/2$} \\
\cline{1-3} \cline{6-8}
\makebox[.5in]{$M_1$} & \makebox[.5in]{$M_2$} & \makebox[.5in]{$M_3$} & & & 
\makebox[.5in]{$M_1$} & \makebox[.5in]{$M_2$} & \makebox[.5in]{$M_3$} \\ 
\cline{1-3} \cline{6-8}
15/2 & 13/2 & $-11/2$ & & & 15/2 & 13/2 & $-13/2$ \\
 & 11/2 & $-9/2$ & & &  & 11/2 & $-11/2$ \\
 & 9/2 & $-7/2$ & & &  & 9/2 & $-9/2$ \\
 & 7/2 & $-5/2$ & & &  & 7/2 & $-7/2$ \\
 & 5/2 & $-3/2$ & & &  & 5/2 & $-5/2$ \\
 & 3/2 & $-1/2$ & & &  & 3/2 & $-3/2$ \\
 & &  & & &  & 1/2 & $-1/2$ & $\longleftarrow$ \\
\cline{1-3} \cline{6-8}
13/2 & 11/2 & $-7/2$ & & & 13/2 & 11/2 & $-9/2$ \\
 & 9/2 & $-5/2$ & & &  & 9/2 & $-7/2$ \\
 & 7/2 & $-3/2$ & & &  & 7/2 & $-5/2$ \\
 & 5/2 & $-1/2$ & & &  & 5/2 & $-3/2$ \\
 & 3/2 & 1/2 & & &  & 3/2 & $-1/2$ \\ 
\cline{1-3} \cline{6-8}
11/2 & 9/2 & $-3/2$ & & & 11/2 & 9/2 & $-5/2$ \\
 & 7/2 & $-1/2$ & & &  & 7/2 & $-3/2$ \\
 & 5/2 & 1/2 & & &  & 5/2 & $-1/2$ \\
 & &  & & &  & 3/2 & 1/2 & $\longleftarrow$\\
\cline{1-3} \cline{6-8}
9/2 & 7/2 & 1/2 & & & 9/2 & 7/2 & $-1/2$ \\
 & 5/2 & 3/2 & & &  & 5/2 & 1/2 \\
\cline{1-3} \cline{6-8}
 & &  & & & 7/2 & 5/2 & 3/2 & $\longleftarrow$
\end{tabular}
\caption{\footnotesize In this example for $j=15/2$, we can see on the 
left-hand side all possible states with $M=j+1=17/2$; while on the right-hand
side, there are all possible states with $M=j=15/2$. Only the states with an
arrow correspond to $J=j$ (see text).}
\label{tab:exam} 
\end{table}

Although it is not necessary for the above derivation, it is illuminating to
show all the states with $J=j+1$ and $M=j+1$, and those with $J=j$, $M=j$. We 
do so in Table~\ref{tab:exam} for $j=15/2$. We see from this table why it is 
simpler to lower $M_3$ for $J=j+1$ rather than, say, $M_1$. For $M_3$ there is
smooth sailing---all lowerings are acceptable. However, if we lower $M_1$, some
of the resulting states will have $M_1 (\rm final)=M_2$, which is not 
acceptable. Indeed, this occurs for the very first state in the table,
($15/2, 13/2, -11/2$); when $M_1$ is lowered, we get ($13/2,13/2,-11/2$).

Following a procedure similar to the one described above, we can also obtain
the number of states with $J=j+1$. We start with states $J=j+2$, $M=j+2$; then
we lower $M_3$ in one unit, and the additional states with $M=j+1$ (which will
have again the form $M_3=M_2-1$) constitute the number of states with $J=j+1$.
For the case of $j=15/2$, these states are two: ($13/2,3/2,1/2$) and 
($9/2,5/2,3/2$). Thus, in general, we have
\begin{equation}
N_{j+1} < \frac{2j+1}{6}, \label{njplus1}
\end{equation}
where $N_{j+1}$ stands for the number of states with $J=j+1$.

It has been noted by Rosensteel and Rowe~\cite{rr03} that the number of $J=0$ 
states for four fermions can be written as
\begin{equation}
\frac{1}{3} \left( \frac{2j+1}{2} + 2\sum_{{\rm even }\,J_0} 
(2J_0+1) \left\{ \matrix{j & j & J_0 \cr j & j & J_0} \right\}
\right) = \left[ \frac{2j+3}{6} \right]. 
\end{equation}
Using this result, Zhao et al.~\cite{zagy03} showed that 
\begin{equation}
{\rm SUM6}j \equiv \sum_{{\rm even}\, J_0} (2J_0+1) 
\left\{ \matrix{j & j & J_0 \cr j & j & J_0} \right\}
\end{equation}
has a modular behaviour. The values are ($-0.5,0.5,0$) for $j$ values
($1/2,3/2,5/2$), and they repeat after that, i.e., they are the same for
($7/2,9/2,11/2$) and for ($13/2,15/2,17/2$), etc.
In one of our previous works~\cite{ze05}, we mention other works by Zhao 
et~al.~\cite{za03,za04}. To this we should add their preprint of 
Ref.~\cite{za05}.

For the case $J=j+1$, we can get an analogous result. We have shown in 
Eq.~(\ref{njplus1}) that the number of $J=j+1$ states is less than $(2j+1)/6$,
which is the same as $[j/3]$. The analogous relation for $J=j+1$ is
\begin{equation}
\frac{1}{3} \left( \frac{2j-1}{2} - 2\sum_{{\rm even}\, J_0} (2J_0+1)
\left\{ \matrix{j & j & J_0 \cr j & j+1 & J_0} \right\} \right) =
\left[ \frac{j}{3} \right].
\end{equation}
From this we find that $\sum_{{\rm even}\, J_0} (2J_0+1) \left\{ 
\matrix{j & j & J_0 \cr j & j+1 & J_0} \right\}=(0,1/2,1),(0,1/2,1)$, etc., for
$j=(1/2,3/2,5/2), (7/2,9/2,11/2)$, etc. This is a special case of a formula
derived by Zhao and Arima~\cite{za04}. The special feature of our work is that
we use a very lowbrow tecnique to get our results.

The sum over all $J_0$ is easier to obtain than that over even $J_0$. Following
the method of Schwinger~\cite{s65}, we find the sum is  unity.

If we consider states with more than 3 particles, we do run into problems where
states with $M=j+1$ can have $M_f=-j$. Consider, for example, 5 neutrons in the
$g_{9/2}$ shell. For this configuration, there are 5 states with $M=j=9/2$, 
where $M_5=M_4-1$. They are as follows

\begin{tabular}{l}
($9/2,7/2,5/2,-5/2,-7/2$) \\
($9/2,7/2,1/2,-3/2,-5/2$) \\
($9/2,5/2,3/2,-3/2,-5/2$) \\
($9/2,3/2,1/2,-1/2,-3/2$) \\
($7/2,5/2,1/2,-1/2,-3/2$) 
\end{tabular}

\noindent However, there are only three $J=j=9/2$ states with the configuration
$(g_{9/2})^5$. The resolution of this dilemma is to notice that for 
$J=j+1=11/2$, there are two states with $M_5=-j=-9/2$. Here we cannot lower 
$M_5$ to reach a state with $J=j=9/2$. these states are

\begin{tabular}{l}
($9/2,7/2,5/2,-1/2,-9/2$) \\
($9/2,7/2,3/2,1/2,-9/2$).
\end{tabular}

\noindent Thus, from the set of 5 states with $M=9/2$ and $M_5=M_4-1$ written
above, two of them still belong to the $J=11/2$ states; and so the number of
$J=9/2$ states is $5-2=3$, and everything is consistent. In general, we could 
say that the number of states with $J=j$ is equal to the number of states with
$M=j$ and $M_f=M_{f-1}-1$, minus the number of $J=j+1$ states with $M_f=-j$.

In the above considerations, we found the work of Bayman and Lande~\cite{bl66}
to be a very useful guide.

\begin{center}
{\Large \bf Acknowledgments}
\end{center}

One of us (L.Z.) is grateful for suppport from INT Seattle in Fall 2004, where 
he found discusions with Igal Talmi and Wick Haxton to be very valuable. This 
work was supported by the U.S. Dept.~of~Energy under Grant 
No.~DE-FG0104ER04-02. A.E. is supported by a grant financed by the 
Secretar\'{\i}a de Estado de Educaci\'on y Universidades (Spain) and cofinanced
by the European Social Fund.

\end{document}